\documentstyle[prb,aps,eqsecnum,preprint,tighten,epsf,floats]{revtex}
\begin{document}
\draft
\title{Berry phase and adiabaticity of a spin diffusing in a non-uniform magnetic field}
\author{S. A. van Langen$^a$, H. P. A. Knops$^a$, J. C. J. Paasschens$^{a,b}$,
and C. W. J. Beenakker$^a$}
\address{$^a$ Instituut-Lorentz, Leiden University,
P.O. Box 9506, 2300 RA Leiden, The Netherlands \\
$^b$ Philips Research Laboratories, 5656 AA Eindhoven, The Netherlands}
\maketitle
\begin{abstract}
An electron spin moving adiabatically in a strong, spatially non-uniform magnetic field 
accumulates a geometric phase or Berry phase, which might be observable as
a conductance oscillation in a mesoscopic ring. Two contradicting theories
exist for how strong the magnetic field should be to ensure adiabaticity
if the motion is diffusive. To resolve this controversy, we study the effect
of a non-uniform magnetic field on the spin polarization and on the weak-localization
effect. The diffusion equation for the Cooperon is solved exactly. Adiabaticity
requires that the spin-precession time is short compared to the 
elastic scattering time --- it is not sufficient that it is short compared to the
diffusion time around the ring. This strong condition severely complicates
the experimental observation.
\smallskip\\
\pacs{PACS: 03.65.Bz, 72.80.Ng, 73.20.Fz, 73.23.-b}
\end{abstract}

\section{Introduction}
The adiabatic theorem of quantum mechanics implies that the final state of a particle
that moves slowly along a closed path is identical to the initial eigenstate --- up to a
phase factor. The Berry phase is a time-independent contribution to this phase,
depending only on the geometry of the path.\cite{berry} A simple example is a spin-$1/2$
in a rotating magnetic field $\bf B$, where the Berry phase equals half the solid angle
swept by $\bf B$. It was proposed by Stern \cite{stern} to measure the Berry phase
in the conductance $G$ of a mesoscopic ring in a spatially rotating magnetic field.
Oscillations of $G$ as a function of the swept solid angle were predicted,
similar to the Aharonov-Bohm oscillations as a function of the enclosed flux.\cite{ab}

An important practical difference between the two effects is that the Aharonov-Bohm
oscillations exist at arbitrarily small magnetic fields, whereas for the oscillations 
due to the Berry phase the magnetic field should be sufficiently strong to allow the
spin to adiabatically follow the changing direction. 
Generally speaking, adiabaticity requires that the precession frequency
$\omega_{\rm B}$ is large compared to the reciprocal of the characteristic timescale
$t_{\rm c}$ on which $\bf B$ changes direction. We know that
$\omega_{\rm B}=g\mu_{\rm B}B/2\hbar$, with $g$ the Land{\'e}-factor and $\mu_{\rm B}$
the Bohr magneton. The question is, what is $t_{\rm c}$? In a ballistic ring there
is only one candidate, the circumference $L$ of the ring divided by the Fermi velocity $v$.
(For simplicity we assume that $L$ is also the scale on which the field direction changes.)
In a diffusive ring there are two candidates: the elastic scattering time $\tau$ and the
diffusion time $\tau_{\rm d}$ around the ring. They differ by a factor
$\tau_{\rm d}/\tau \simeq (L/\ell)^2$, where $\ell=v\tau$ is the mean free path.
Since, by definition, $L\gg\ell$ in a diffusive system, the two time scales are
far apart. Which of the two time scales is the relevant one is still under
debate.\cite{sternrev}

Stern's original proposal \cite{stern} was that
\begin{equation}
\label{critstern}
\omega_{\rm B}\gg \frac{1}{\tau}
\end{equation}
is necessary to observe the Berry-phase oscillations. For realistic values of $g$
this requires magnetic fields in the quantum Hall regime, outside the range of
validity of the semiclassical theory. We call Eq.\ (\ref{critstern}) the
``pessimistic criterion''.
In a later work, \cite{lsg} Loss, Schoeller, and Goldbart (LSG) concluded that adiabaticity is
reached already at much weaker magnetic fields, when
\begin{equation}
\label{critloss}
\omega_{\rm B}\gg\frac{1}{\tau_{\rm d}}\simeq\frac{1}{\tau}\left(\frac{\ell}{L}\right)^2.
\end{equation}
This ``optimistic criterion'' has motivated experimentalists to search for the Berry-phase
oscillations in disordered conductors, \cite{experiments} and was invoked in a recent
study of the conductivity of mesoscopic ferromagnets.\cite{lyanda}
In this paper, we re-examine the semiclassical theory of LSG to resolve the controversy.

The Berry-phase oscillations in the conductance result from a periodic modulation of the
weak-localization correction, and require the solution of a diffusion equation
for the Cooperon propagator. To solve this problem we need to consider
the coupled dynamics of four spin-degrees of freedom. (The Cooperon has four spin indices.)
To gain insight we first examine in Sec.\ \ref{transmission} the simpler problem of the
dynamics of a single spin variable, by studying the randomization of a spin-polarized 
electron gas by a non-uniform magnetic field. We start at the level of the Boltzmann
equation and then make the diffusion approximation. We show how the diffusion
equation can be solved exactly for the first two moments of the polarization.
The same procedure is used in Sec.\ \ref{localization} to arrive at a diffusion
equation for the Cooperon. This equation coincides with the equation derived
by LSG in the weak-field regime $\omega_{\rm B}\tau\ll 1$, but is different in the
strong-field regime $\omega_{\rm B}\tau\gtrsim 1$. We present an exact solution for
the weak-localization correction and compare with the findings of LSG.

Our conclusion both for the polarization and for the weak-localization correction is
that adiabaticity requires $\omega_{\rm B}\tau\gg 1$. Regrettably, the pessimistic
criterion (\ref{critstern}) is correct, in agreement with Stern's original conclusion.
The optimistic criterion (\ref{critloss}) advocated by LSG turns out to be the criterion
for maximal randomization of the spin by the magnetic field, and not the criterion for
adiabaticity.

\section{Spin-resolved transmission}
\label{transmission}
\subsection{Formulation of the problem}
Consider a conductor in a magnetic field $\bf B$, containing a disordered
segment (length $L$, mean free path $\ell$ at Fermi velocity $v$) in which the
magnetic field changes its direction. An electron at the Fermi level with spin
up (relative to the local magnetic field) is injected at one end and reaches
the other end. What is the probability that its spin is up?

\begin{figure}[ht]
\epsfxsize=0.7\hsize
\hspace*{\fill}
\vspace*{-0ex}\epsffile{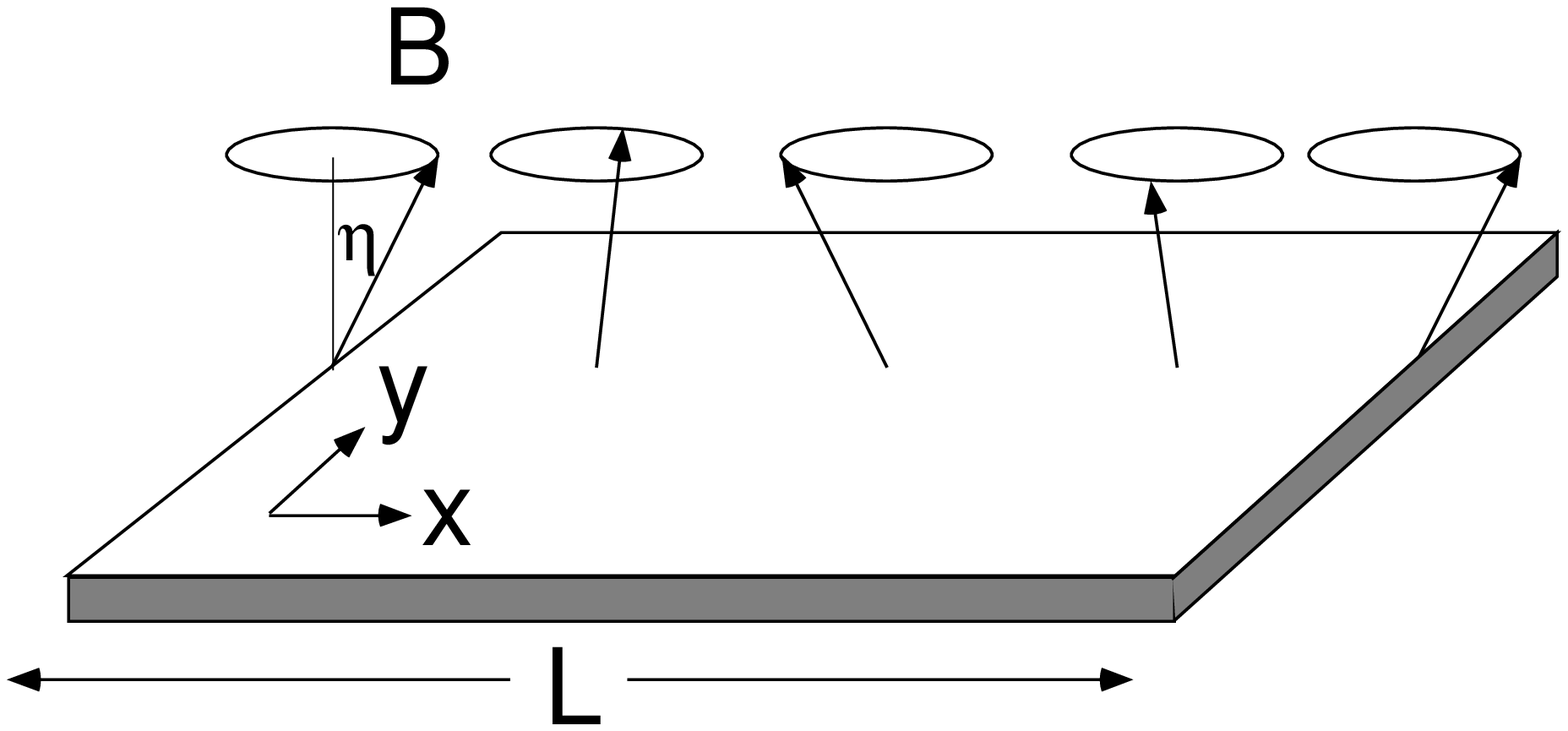}\vspace*{0ex}
\hspace*{\fill}
\medskip
\caption[]{Schematic drawing of a two-dimensional electron gas in the spatially
rotating magnetic field of Eq.\ (\ref{field}), with $f=1$.}
\label{fig1}
\end{figure}

For simplicity we take for the conductor a two-dimensional electron gas (in the  $x$-$y$ plane,
with the disordered region between $x=0$ and $x=L$), and we ignore the 
curvature of the electron trajectories by the Lorentz force. The problem becomes
effectively one-dimensional by assuming that $\bf B$ depends on $x$ only.
We choose a rotation of $\bf B$ in the $x$-$y$-plane, according to 
\begin{equation}
\label{field}
{\bf B}(x,y,z=0)=
\left(B\sin\eta\cos\case{2\pi fx}{L},B\sin\eta\sin\case{2\pi fx}{L},B\cos\eta\right),
\end{equation}
with $\eta$ and $f$ arbitrary parameters. The geometry is sketched in
Fig.\ \ref{fig1}. We treat the orbital motion semiclassically, within
the framework of the Boltzmann equation. (This is justified if the Fermi wavelength
is much smaller than $\ell$.) The spin dynamics requires a fully quantum mechanical
treatment. We assume that the Zeeman energy $g\mu_{\rm B}B$ is much smaller than the
Fermi energy $\frac{1}{2}mv^2$, so that the orbital motion is independent
of the spin.

We introduce the probability density $P(x,\phi,\xi,t)$ for the electron to be
at time $t$ at position $x$ with velocity ${\bf v}=(v\cos\phi,v\sin\phi,0)$, in
the spin state with spinor $\xi =(\xi_1,\xi_2)$. The dynamics of $\xi$ depends
on the local magnetic field according to
\begin{equation}
\label{schroedinger}
\frac{d\xi}{dt}=\frac{ig\mu_{\rm B}}{2\hbar}{\bf B}\cdot\mbox{\boldmath$\sigma$}\,\xi,
\end{equation}
where $\mbox{\boldmath$\sigma$}=(\sigma_x,\sigma_y,\sigma_z)$ is the vector of Pauli matrices.
It is convenient to decompose $\xi=\chi_1\xi_\uparrow +\chi_2\xi_\downarrow$ into
the local eigenstates $\xi_\uparrow,\xi_\downarrow$ of ${\bf B}\cdot\mbox{\boldmath$\sigma$}$,
\begin{mathletters}
\label{localbasis}
\begin{equation}
\xi_\uparrow =
\pmatrix{\cos\frac{\eta}{2}\,{\rm e}^{-i\pi fx/L}  \cr
         \sin\frac{\eta}{2}\,{\rm e}^{i\pi fx/L}},~~~~
\xi_\downarrow =
\pmatrix{-\sin\frac{\eta}{2}\,{\rm e}^{-i\pi fx/L}  \cr
         \cos\frac{\eta}{2}\,{\rm e}^{i\pi fx/L}},
\end{equation}
\begin{equation}
{\bf B}\cdot\mbox{\boldmath$\sigma$}\,\xi_\uparrow = B\xi_\uparrow,~~~~
{\bf B}\cdot\mbox{\boldmath$\sigma$}\,\xi_\downarrow = -B\xi_\downarrow,
\end{equation}
\end{mathletters}
and use the real and imaginary parts of the coefficients $\chi_1,\chi_2$ as
variables in the Boltzmann equation. The dynamics of the vector of coefficients
$c=(c_1,c_2,c_3,c_4)=({\rm Re}\,\chi_1,{\rm Im}\,\chi_1,{\rm Re}\,\chi_2,{\rm Im}\,\chi_2)$
is given by
\begin{mathletters}
\begin{equation}
\frac{dc}{dt} = \frac{1}{\tau} M c,~~M=M_0+M_1\cos\phi,
\end{equation}
\begin{equation}
\label{defm}
M_0 = \omega_{\rm B}\tau \pmatrix{
0&-1&0&0 \cr 1&0&0&0 \cr 0&0&0&1 \cr 0&0&-1&0 },~~~~~
M_1=\frac{\pi f\ell}{L} \pmatrix{
0          & -\cos\eta  & 0        &   \sin\eta  \cr
 \cos\eta  & 0         & -\sin\eta &   0          \cr
0          & \sin\eta & 0        &   \cos\eta   \cr
-\sin\eta   & 0         & -\cos\eta &   0   
},
\end{equation}
\end{mathletters}
where $\omega_{\rm B}=g\mu_{\rm B}B/2\hbar$ is the precession frequency of the spin.
The Boltzmann equation takes the form 
\begin{equation}
\label{boltzmann}
\tau\frac{\partial}{\partial t} P(x,\phi,c,t) =
-\ell\cos\phi\frac{\partial P}{\partial x}
-\sum_{i,j}\frac{\partial}{\partial c_i} \left(M_{ij}c_j P\right)
-P + \int_0^{2\pi}\frac{d\phi'}{2\pi} P(x,\phi',c,t),
\end{equation}
where we have assumed isotropic scattering (rate $1/\tau=v/\ell$).

We look for a stationary solution to the Boltzmann equation, so the left-hand-side of 
Eq.\ (\ref{boltzmann}) is zero and we omit the argument $t$ of $P$.
A stationary flux of particles with an isotropic velocity distribution
is injected at $x=0$, their spins all aligned with the local magnetic field
(so $\chi_2=0$ at $x=0$). Without loss of generality we may assume that
$\chi_1=1$ at $x=0$. No particles are incident from the other end, at $x=L$.
Thus the boundary conditions are
\begin{mathletters}
\label{beecee}
\begin{eqnarray}
\label{beginwire}
&& P(x=0,\phi,c)=\delta (c_1-1)\delta (c_2)\delta (c_3)\delta (c_4)~{\rm if}~\cos\phi > 0, \\
\label{endwire}
&& P(x=L,\phi,c) = 0~{\rm if}~\cos\phi < 0.
\end{eqnarray}
\end{mathletters}

This completes the formulation of the problem. We compare two methods of solution.
The first is an exact numerical solution of the Boltzmann equation using the
Monte Carlo method. The second is an approximate analytical solution using the
diffusion approximation, valid for $L\gg \ell$. We begin with the latter.

\subsection{Diffusion approximation}
The diffusion approximation amounts to the assumption that $P$ has a simple
cosine-dependence on $\phi$,
\begin{equation}
\label{diffansatz}
P(x,\phi,c)=N(x,c)+J(x,c)\cos\phi.
\end{equation}
To determine the density $N$ and current $J$ we 
substitute Eq.\ (\ref{diffansatz}) into Eq.\ (\ref{boltzmann}) and integrate
over $\phi$. This gives
\begin{eqnarray}
\ell\frac{\partial J}{\partial x} &=&
-\frac{\partial}{\partial c}\left( 2 M_0 c N+M_1 c J\right).
\label{diffpde1}
\end{eqnarray}
Similarly, multiplication with $\cos\phi$ before integration gives 
\begin{eqnarray}
\ell\frac{\partial N}{\partial x} &=&
-\frac{\partial}{\partial c} \left( M_0 c J+ M_1 c N\right) - J.
\label{diffpde2}
\end{eqnarray}
Thus we have a closed set of partial differential equations for the
unknown functions $N(x, c)$ and $J(x, c)$. Boundary conditions are obtained by
multiplying Eq.\ (\ref{beecee}) with $\cos\phi$ and integrating over $\phi$:
\begin{mathletters}
\label{diffbc}
\begin{eqnarray}
&& N(x=0,c)+\frac{\pi}{4}J(x=0,c)=\delta (c_1-1)\delta (c_2)\delta (c_3)\delta (c_4),\\
&& N(x=L,c)-\frac{\pi}{4}J(x=L,c)=0.
\end{eqnarray}
\end{mathletters}

We seek the spin polarization $p=c_1^2+c_2^2-c_3^2-c_4^2$ of the transmitted 
electrons, characterized by the distribution
\begin{equation}
\label{goal}
P(p)=\frac{\int\!dc\,J(x=L,c)\delta(c_1^2+c_2^2-c_3^2-c_4^2-p)}{\int\!dc\, J(x=L,c)}. 
\end{equation}
(The notation $\int\!dc\,\equiv\int\!dc_1\,\int\!dc_2\,\int\!dc_3\,\int\!dc_4$
indicates an integration over the spin variables.) We compute the first two moments
of $P(p)$.
The first moment $\overline{p}$ is the fraction of transmitted electrons with
spin up minus the fraction with spin down, averaged quantum mechanically over the
spin state and statistically over the disorder.
The variance Var $p=\overline{p^2}-\overline{p}^2$ gives an
indication of the magnitude of the statistical fluctuations.

Integration of Eqs.\ (\ref{diffpde1})--(\ref{diffbc}) over the spin variables
yields the equations and boundary conditions for the functions
$N(x)=\int\!dc\,N(x,c)$ and $J(x)=\int\!dc\,J(x,c)$:
\begin{mathletters}
\label{momzero}
\begin{eqnarray}
&&\ell\frac{dN}{dx}=-J,~~\frac{dJ}{dx} = 0, \label{momzerodiff} \\
&& N(0)+\frac{\pi}{4}J(0)=1,~~N(L)-\frac{\pi}{4}J(L)=0.
\end{eqnarray}
\end{mathletters}
The solution
\begin{equation}
\label{solnorm}
J(x) =  \left( \frac{\pi}{2} + \frac{L}{\ell} \right)^{-1}
\end{equation}
determines the denominator of Eq.\ (\ref{goal}).

To determine $\overline{p}$ we multiply Eqs.\ (\ref{diffpde1}) and (\ref{diffpde2})
with $\chi_\alpha\chi_\beta^\ast$ and integrate over $c$. (Recall that
$\chi_1=c_1+ic_2,\chi_2=c_3+ic_4$.)
It follows upon partial integration that
\begin{mathletters}
\begin{eqnarray}
&&\int\! dc\, \chi_\alpha\chi_\beta^\ast \frac{\partial}{\partial c}\left(M_0 c f\right)=
-\sum_{\rho,\sigma}
\left(S_{\alpha\rho}\delta_{\beta\sigma}-\delta_{\alpha\rho}S_{\beta\sigma}\right)
\int\! dc\, \chi_\rho \chi_\sigma^\ast f, \\
&&\int\! dc\, \chi_\alpha\chi_\beta^\ast \frac{\partial}{\partial c}\left(M_1 c f\right) =
-\sum_{\rho,\sigma}
\left(T_{\alpha\rho}\delta_{\beta\sigma}-\delta_{\alpha\rho}T_{\beta\sigma}\right)
\int\! dc\, \chi_\rho \chi_\sigma^\ast f,
\end{eqnarray}
\end{mathletters}
for arbitrary functions $f(x,c)$. The $2\times 2$ matrices $S,T$ are defined by
\begin{equation}
S=i\omega_{\rm B}\tau\sigma_z,
~~T=\frac{i\pi f\ell}{L}\left(\sigma_z\cos\eta-\sigma_x\sin\eta\right). 
\end{equation}
In this way we find that the moments
\begin{mathletters}
\begin{eqnarray}
N_{\alpha\beta}(x)&=&\int\!dc\,\chi_\alpha\chi^\ast_\beta N(x,c), \\
J_{\alpha\beta}(x)&=&\int\!dc\,\chi_\alpha\chi^\ast_\beta J(x,c),
\end{eqnarray}
\end{mathletters}
satisfy the ordinary differential equations
\begin{mathletters}
\label{momdiff}
\begin{eqnarray}
\label{momdiffa}
&& \ell\frac{dN_{\alpha\beta}}{dx} = \sum_{\rho,\sigma}
\left(T_{\alpha\rho}\delta_{\beta\sigma}-\delta_{\alpha\rho}T_{\beta\sigma}\right)
N_{\rho\sigma}
+\sum_{\rho,\sigma}
\left(S_{\alpha\rho}\delta_{\beta\sigma}-\delta_{\alpha\rho}S_{\beta\sigma}\right)
J_{\rho\sigma} -J_{\alpha\beta}, \\
&& \ell\frac{dJ_{\alpha\beta}}{dx} =
2\sum_{\rho,\sigma}
\left(S_{\alpha\rho}\delta_{\beta\sigma}-\delta_{\alpha\rho}S_{\beta\sigma}\right)
N_{\rho\sigma}
+\sum_{\rho,\sigma}
\left(T_{\alpha\rho}\delta_{\beta\sigma}-\delta_{\alpha\rho}T_{\beta\sigma}\right)
J_{\rho\sigma},
\end{eqnarray}
\end{mathletters}
with boundary conditions
\begin{mathletters}
\begin{eqnarray}
&& N_{\alpha\beta}(x=0)+\frac{\pi}{4}J_{\alpha\beta}(x=0)=\delta_{\alpha 1}\delta_{\beta 1}, \\
&& N_{\alpha\beta}(x=L)-\frac{\pi}{4}J_{\alpha\beta}(x=L)= 0.
\end{eqnarray}
\end{mathletters}
The mean polarization $\overline{p}$ is determined by $J_{\alpha\beta}$ according to
\begin{equation}
\label{avpdef}
\overline{p}=\frac{J_{11}(L)-J_{22}(L)}{J(L)}=
\left(\frac{\pi}{2}+\frac{L}{\ell}\right)\left[J_{11}(L)-J_{22}(L)\right].
\end{equation}

\begin{figure}[ht]
\epsfxsize=0.6\hsize
\hspace*{\fill}
\vspace*{-0ex}\epsffile{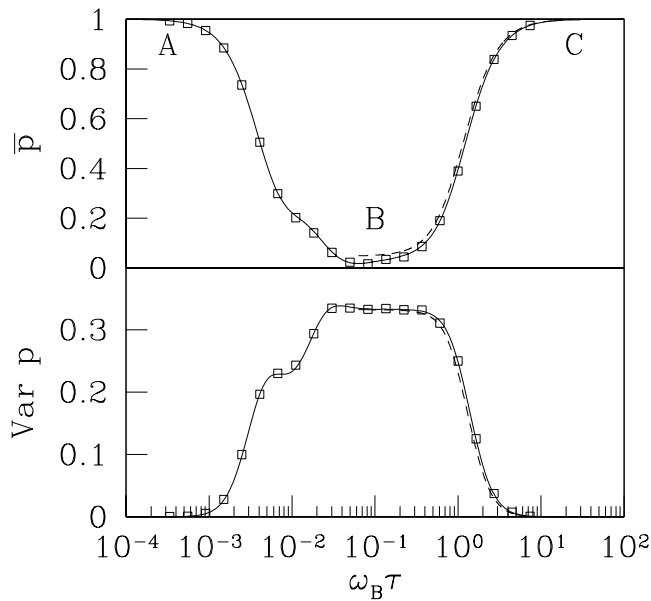}\vspace*{0ex}
\hspace*{\fill}
\medskip
\caption[]{Average and variance of the spin polarization $p$ of the current
transmitted through a two-dimensional region of length $L=25\,\ell$,
as a function of $\omega_{\rm B}\tau$, for a magnetic field given by Eq.\ (\ref{field})
with $\eta=\pi/3$ and $f=1$.
The data points result from Monte Carlo simulations of the Boltzmann
equation (\ref{boltzmann}),
the solid curves result from the diffusion approximation (\ref{diffansatz}),
and the dashed curves are the asymptotic formulas (\ref{avp}) and (\ref{varp}).
Notice the transient regime (A), the randomized regime (B), and the adiabatic
regime (C). }
\label{fig2}
\end{figure}

Since Eq.\ (\ref{momdiff}) is linear in the $8$ functions
$N_{\alpha\beta}(x),J_{\alpha\beta}(x)$ ($\alpha,\beta=1,2$), a solution requires 
the eigenvalues and right eigenvectors of the $8\times 8$ matrix of coefficients.
These can be readily computed numerically for any values of $L/\ell$ and
$\omega_{\rm B}\tau$. We have found an analytic asymptotic solution for $L/\ell\gg 1$
and $\omega_{\rm B}\tau\gg (f\ell/L)^2$, given by 
\begin{equation}
\label{avp}
\overline{p}=\frac{k}{\sinh k},~~~k=\frac{2\pi f\sin\eta}{\sqrt{1+(2\omega_{\rm B}\tau)^2}}.
\end{equation}
In Fig.\ \ref{fig2} we compare the numerical solution (solid curve) with Eq.\ (\ref{avp})
(dashed curve) for $L/\ell=25$ and $\eta=\pi/3,f=1$. The two curves are almost
indistinguishable, except for the smallest values of $\omega_{\rm B}\tau$. 

In a similar way, we compute the second moment of $P(p)$ by multiplying
Eqs.\ (\ref{diffpde1}) and (\ref{diffpde2})
with $\chi_\alpha\chi_\beta^\ast\chi_\gamma\chi_\delta^\ast$ and integrating over
$c$. The result is a closed set of equations
\begin{mathletters}
\label{vardiff}
\begin{eqnarray}
\label{vardiffa} 
&& \ell \frac{d}{dx} N_{\alpha\beta\gamma\delta} = \sum_{\mu,\nu,\rho,\sigma} \left(
L^{\mu\nu\rho\sigma}_{\alpha\beta\gamma\delta} N_{\mu\nu\rho\sigma}
+ K^{\mu\nu\rho\sigma}_{\alpha\beta\gamma\delta} J_{\mu\nu\rho\sigma} \right)
- J_{\alpha\beta\gamma\delta}, \\
&& \ell \frac{d}{dx} J_{\alpha\beta\gamma\delta} = \sum_{\mu,\nu,\rho,\sigma} \left(
2K^{\mu\nu\rho\sigma}_{\alpha\beta\gamma\delta} N_{\mu\nu\rho\sigma}
+ L^{\mu\nu\rho\sigma}_{\alpha\beta\gamma\delta} J_{\mu\nu\rho\sigma} \right) ,
\end{eqnarray}
\end{mathletters}
where we have defined 
\begin{mathletters}
\begin{eqnarray} 
&& K_{\alpha\beta\gamma\delta}^{\mu\nu\rho\sigma} = 
S_{\alpha\mu}\delta_{\beta\nu}\delta_{\gamma\rho}\delta_{\delta\sigma}
-\delta_{\alpha\mu}S_{\beta\nu}\delta_{\gamma\rho}\delta_{\delta\sigma}
+\delta_{\alpha\mu}\delta_{\beta\nu}S_{\gamma\rho}\delta_{\delta\sigma}
-\delta_{\alpha\mu}\delta_{\beta\nu}\delta_{\gamma\rho}S_{\delta\sigma}, \\
&& L_{\alpha\beta\gamma\delta}^{\mu\nu\rho\sigma} = 
T_{\alpha\mu}\delta_{\beta\nu}\delta_{\gamma\rho}\delta_{\delta\sigma}
-\delta_{\alpha\mu}T_{\beta\nu}\delta_{\gamma\rho}\delta_{\delta\sigma}
+\delta_{\alpha\mu}\delta_{\beta\nu}T_{\gamma\rho}\delta_{\delta\sigma}
-\delta_{\alpha\mu}\delta_{\beta\nu}\delta_{\gamma\rho}T_{\delta\sigma},
\end{eqnarray}
\end{mathletters}
\begin{mathletters}
\begin{eqnarray}
N_{\alpha\beta\gamma\delta}(x) &=& \int\!dc\,
\chi_\alpha\chi^\ast_\beta\chi_\gamma\chi^\ast_\delta N(x,c), \\ 
J_{\alpha\beta\gamma\delta}(x) &=& \int\!dc\,
\chi_\alpha\chi^\ast_\beta\chi_\gamma\chi^\ast_\delta J(x,c).
\end{eqnarray}
\end{mathletters}
The boundary conditions on the functions $N_{\alpha\beta\gamma\delta}$ and 
$J_{\alpha\beta\gamma\delta}$ are 
\begin{eqnarray}
&& N_{\alpha\beta\gamma\delta}(x=0)+\frac{\pi}{4}J_{\alpha\beta\gamma\delta}(x=0)
=\delta_{\alpha 1}\delta_{\beta 1}\delta_{\gamma 1}\delta_{\delta 1}, \\
&& N_{\alpha\beta\gamma\delta}(x=L)-\frac{\pi}{4}J_{\alpha\beta\gamma\delta}(x=L)= 0.
\end{eqnarray}
The second moment $\overline{p^2}$ is determined by
\begin{equation}
\overline{p^2}=\left(\frac{\pi}{2}+\frac{L}{\ell}\right)
\left[ J_{1111}(x=L)-J_{1122}(x=L)-J_{2211}(x=L)+J_{2222}(x=L) \right].
\end{equation}
The numerical solution is plotted also in Fig.\ \ref{fig2}, together with the 
asymptotic expression 
\begin{equation}
\label{varp}
{\rm Var}\, p = \frac{1}{3}+\frac{2k\sqrt{3}}{3\sinh\left(k\sqrt{3}\right)}
-\frac{k^2}{\sinh^2 k}.
\end{equation}

It is evident from Eqs.\ (\ref{avp}) and (\ref{varp}), and from Fig.\ \ref{fig2},
that the regime with $\overline{p}=1$, ${\rm Var}\,p=0$ is entered for
$\omega_{\rm B}\tau \gtrsim f$ [for $\sin\eta ={\cal O} (1)$], in agreement with Stern's
criterion (\ref{critstern}) for adiabaticity. For smaller $\omega_{\rm B}\tau$ 
adiabaticity is lost. There is a transient regime
$\omega_{\rm B}\tau \ll (f\ell/L)^2$, in which the precession frequency is
so low that the spin remains in the same state during the entire diffusion process.
For $(f\ell/L)^2 \ll \omega_{\rm B}\tau \ll f$ the average polarization reaches a
plateau value close to zero with a finite variance. For a sufficiently non-uniform
field, $f\sin\eta\gg 1$, we find in this regime $\overline{p}=0$ and ${\rm Var}\, p=1/3$,
which means that the spin state is completely randomized. The transient regime,
the randomized regime, and the adiabatic regime are indicated in Fig.\ \ref{fig2}
by the letters A, B, and C.

\subsection{Comparison with Monte Carlo simulations}
In order to check the diffusion approximation we solved the full Boltzmann equation
by means of a Monte Carlo simulation. A particle is moved from $x=0$ over a distance
$\ell_1$ in the direction $\phi_1$, then over a distance $\ell_2$ in the direction $\phi_2$,
and so on, until it is reflected back to $x=0$ or transmitted to $x=L$.
The step lengths $\ell_i$ are chosen randomly from a Poisson distribution with
mean $\ell$. The directions $\phi_i$ are chosen uniformly from $[0,2\pi]$, except for
the initial direction $\phi_1$, which is distributed $\propto\cos\phi_1$.
The spin components are given by
\begin{equation}
\pmatrix{\chi_1 \cr \chi_2} =
\prod_i {\rm e}^{\left(S + T \cos\phi_i \right)\ell_i /\ell}
\pmatrix{ 1 \cr 0 }.
\end{equation}
To find $\overline{p^n}$, one has to average $\left(|\chi_1|^2-|\chi_2|^2\right)^n$
over the transmitted particles.
The results for $L/\ell=25$ are shown in Fig. \ref{fig2} (data points). They agree
very well with the results of the previous subsection, thus confirming the validity of
the diffusion approximation for $L/\ell\gg 1$.

\section{Weak localization}
\label{localization}

\subsection{Formulation of the problem}
We turn to the effect of the non-uniform magnetic field on the weak-localization
correction of a multiply-connected system. We consider the same geometry as in
Fig.\ \ref{fig1}, but now with periodic boundary conditions --- to model a ring of
circumference $L$. Only the effects of the magnetic field on the spin are included,
to isolate the Berry phase from the conventional Aharonov-Bohm phase. As in the 
previous subsection, we assume that the orbital motion is independent of the spin dynamics.
We follow LSG in applying the semiclassical theory of Chakravarty and Schmidt
\cite{chakra} to the problem, however, we start at the level of the Boltzmann
equation --- rather than at the level of the diffusion equation --- and make the 
diffusion approximation at a later stage of the calculation.

The weak-localization correction $\Delta G$ to the conductance is given by
\begin{equation}
\label{wl1}
\Delta G = - \frac{e^2D}{\pi\hbar L} \int_0^\infty\!\!dt\, {\rm e}^{-t/\tau_\varphi} C(t),
\end{equation}
where $\tau_\varphi$ is the phase coherence time and the diffusion coefficient $D=vl/d$
in $d$ dimensions. (In our geometry $d=2$.) The ``return quasi-probability'' $C(t)$
is expressed as
a sum over ``Boltzmannian walks'' ${\bf R}(t)$ with ${\bf R}(0)={\bf R}(t)$, 
\begin{equation}
\label{returnprob}
C(t)= \sum_{\left\{{\bf R}(t)\right\}} W\, {\rm Tr}\, (U^+U^-).
\end{equation}
Here $W[{\bf R}(t)]$ is the weight of the Boltzmannian walk for a spinless particle.
The $2\times 2$ matrices $U^\pm[{\bf R}(t)]$ are defined by
\begin{equation}
U^\pm = {\cal T}\exp\left\{\pm\frac{ig\mu_{\rm B}}{2\hbar}\int_0^t\!dt'\,
{\bf B}\biglb({\bf R}(t')\bigrb)
\cdot \mbox{\boldmath$\sigma$}\right\},
\end{equation}
where $\cal T$ denotes a time ordering. The factor ${\rm Tr}\, (U^+U^-)$ in
Eq.\ (\ref{returnprob}) accounts for the phase difference of time-reversed paths.

The Cooperon can be written in terms of a propagator $\chi$,
\begin{equation}
\label{cooperon}
C(t)=\frac{1}{2\pi}\int_0^{2\pi}\!\! d\phi \int_0^{2\pi}\!\! d\phi_{\rm i}\sum_{\alpha,\beta}
\chi_{\alpha\beta\beta\alpha} (x_{\rm i},x_{\rm i};\phi,\phi_{\rm i};t),
\end{equation}
that satisfies the kinetic equation
\begin{eqnarray}
\label{kinetic}
 \left( \frac{\partial}{\partial t} + {\cal B} \right)
\chi_{\alpha\beta\gamma\delta} (x,x_{\rm i};\phi,\phi_{\rm i};t) -
\frac{ig\mu_{\rm B}}{2\hbar} \sum_{\alpha',\gamma'}
\left[
\biglb({\bf B}(x)\cdot\mbox{\boldmath$\sigma$}\bigrb)_{\alpha\alpha'}\delta_{\gamma\gamma'}-
\delta_{\alpha\alpha'}\biglb({\bf B}(x)\cdot\mbox{\boldmath$\sigma$}\bigrb)_{\gamma\gamma'}
\right]
\chi_{\alpha'\beta\gamma'\delta} \nonumber \\
= \delta (t) \delta (x-x_{\rm i}) \delta (\phi-\phi_{\rm i})
\delta_{\alpha\beta} \delta_{\gamma\delta}.
\end{eqnarray}
The Boltzmann operator $\cal B$ is given by
\begin{equation}
{\cal B} = v\cos\phi\frac{\partial}{\partial x}
+ \frac{1}{\tau} - \frac{1}{\tau}\int_0^{2\pi}\!\frac{d\phi}{2\pi}.
\end{equation}

The propagator $\chi$ is a moment of the probability distribution
$P(x,\phi,U^+,U^-,t)$, 
\begin{equation}
\label{propagator}
\chi_{\alpha\beta\gamma\delta} =\int\! dU^+ \int\! dU^- \,
U^+_{\alpha\beta} U^-_{\gamma\delta} P,
\end{equation}
that satisfies the Boltzmann equation
\begin{equation}
\label{boeq}
\left[\frac{\partial}{\partial t} + {\cal B}
+\frac{\partial}{\partial U^+} \left(\frac{dU^+}{dt}\right)
+\frac{\partial}{\partial U^-} \left(\frac{dU^-}{dt}\right)
\right] P(x,\phi,U^+,U^-,t) = 0,
\end{equation}
with initial condition
\begin{equation}
P(x,\phi,U^+,U^-,0) = \delta (x-x_{\rm i}) \delta (\phi-\phi_{\rm i}) \delta (U^+-\openone )
\delta (U^--\openone ).
\end{equation}
The notation $dU^+$ or $dU^-$ indicates the differential of the real and imaginary
parts of the elements of the $2\times 2$ matrix $U^+$ or $U^-$. We will write this
in a more explicit way in the next subsection.

The Boltzmann equation (\ref{boeq}) has the same form as that which we studied in
Sec.\ \ref{transmission}. The difference is that we have four times as many internal
degrees of freedom. Instead of a single spinor $\xi$ we now have two
spinor matrices $U^+$ and $U^-$. A first doubling of the number of degrees of freedom
occurs because we have to follow the evolution of both spin up and spin down. A second
doubling occurs because we have to follow both the normal and the time-reversed 
evolution.

\subsection{Diffusion approximation.}

We make the diffusion approximation to the Boltzmann equation (\ref{boeq}), by
following the steps outlined in Sec.\ \ref{transmission}.
The $4 \times 2$ matrix $u^\pm$ containing the real and imaginary parts of $U^\pm$,
\begin{equation}
u^\pm = \pmatrix{
{\rm Re}\, U^\pm_{11} & {\rm Re}\, U^\pm_{12} \cr
{\rm Im}\, U^\pm_{11} & {\rm Im}\, U^\pm_{12} \cr
{\rm Re}\, U^\pm_{21} & {\rm Re}\, U^\pm_{22} \cr
{\rm Im}\, U^\pm_{21} & {\rm Im}\, U^\pm_{22} },
\end{equation}
has a time evolution governed by
\begin{mathletters}
\begin{equation}
\tau\frac{du^\pm}{dt}= \pm Z(x) u^\pm,
\end{equation}
\begin{equation}
Z(x)=\omega_{\rm B}\tau \pmatrix{
0    &  -\cos\eta   & \sin\eta\sin\frac{2\pi fx}{L} & -\sin\eta\cos\frac{2\pi fx}{L} \cr
\cos\eta  & 0 & \sin\eta\cos\frac{2\pi fx}{L} & \hphantom{-}\sin\eta\sin\frac{2\pi fx}{L} \cr
-\sin\eta\sin\frac{2\pi fx}{L} & -\sin\eta\cos\frac{2\pi fx}{L} & 0 & \cos\eta \cr
\hphantom{-}\sin\eta\cos\frac{2\pi fx}{L} & -\sin\eta\sin\frac{2\pi fx}{L} & -\cos\eta & 0 
}.
\end{equation}
\end{mathletters}
The Boltzmann equation (\ref{boeq}) becomes, in a more explicit notation,
\begin{eqnarray}
\label{freqbe}
\tau \frac{\partial}{\partial t} P(x,\phi,u^+,u^-,t) &=&
-\ell\cos\phi\frac{\partial P}{\partial x}
- \sum_{i,j,k} \frac{\partial}{\partial u^+_{ij}} Z_{ik}(x) u^+_{kj} P
+ \sum_{i,j,k} \frac{\partial}{\partial u^-_{ij}} Z_{ik}(x) u^-_{kj} P \nonumber \\
&& - P + \int_0^{2\pi}\! \frac{d\phi'}{2\pi} P(x,\phi',u^+,u^-,t).
\end{eqnarray}

We now make the diffusion ansatz in the form
\begin{equation}
\int_0^\infty\! dt\, {\rm e}^{-t/\tau_\varphi}\int_0^{2\pi}\! d\phi_{\rm i}\, P = N + J\cos\phi.
\end{equation}
By integrating the Boltzmann equation over $\phi$, once
with weight $1$ and once with weight $\cos\phi$, we obtain two coupled equations for
the functions $N(x,u^+,u^-)$ and $J(x,u^+,u^-)$. Next we multiply both equations with
$U^+_{\alpha\beta}U^-_{\gamma\delta}$ and integrate over the real and imaginary parts of
the matrix elements. The moments $N_{\alpha\beta\gamma\delta}$
and $J_{\alpha\beta\gamma\delta}$ defined by
\begin{mathletters}
\begin{eqnarray}
\label{wlmom}
N_{\alpha\beta\gamma\delta} (x) &=&
\int\! dU^+\int\! dU^-\, U^+_{\alpha\beta} U^-_{\gamma\delta} N, \\
J_{\alpha\beta\gamma\delta} (x) &=&
\int\! dU^+\int\! dU^-\, U^+_{\alpha\beta} U^-_{\gamma\delta} J,
\end{eqnarray}
\end{mathletters}
are found to obey the ordinary differential equations
\begin{mathletters}
\label{odiff}
\begin{eqnarray}
\label{coopdiffa}
\ell\frac{dN_{\alpha\beta\gamma\delta}}{dx} &=&
\frac{ig\mu_{\rm B}\tau}{2\hbar} \sum_{\alpha',\gamma'}
\left[\biglb({\bf B}(x)\cdot\mbox{\boldmath$\sigma$}\bigrb)_{\alpha\alpha'}
\delta_{\gamma\gamma'}-
\delta_{\alpha\alpha'}
\biglb({\bf B}(x)\cdot\mbox{\boldmath$\sigma$} \bigrb)_{\gamma\gamma'}\right]
J_{\alpha'\beta\gamma'\delta} \nonumber \\
&& -\left(1+\tau/\tau_\varphi\right)J_{\alpha\beta\gamma\delta}, \\
\label{coopdiffb}
\ell\frac{dJ_{\alpha\beta\gamma\delta}}{dx} &=&
\frac{ig\mu_{\rm B}\tau}{\hbar} \sum_{\alpha',\gamma'}
\left[\biglb({\bf B}(x)\cdot\mbox{\boldmath$\sigma$}\bigrb)_{\alpha\alpha'}\delta_{\gamma\gamma'}-
\delta_{\alpha\alpha'} \biglb({\bf B}(x)\cdot \mbox{\boldmath$\sigma$}\bigrb)_{\gamma\gamma'}
\right] N_{\alpha'\beta\gamma'\delta} \nonumber \\
&& -(2\tau/\tau_\varphi) N_{\alpha\beta\gamma\delta}
+2\tau\delta_{\alpha\beta}\delta_{\gamma\delta}\delta (x-x_{\rm i}).
\end{eqnarray}
\end{mathletters}
The periodic boundary conditions are
\begin{equation}
\label{ringbc}
N_{\alpha\beta\gamma\delta}(0) = N_{\alpha\beta\gamma\delta}(L),~~~
J_{\alpha\beta\gamma\delta}(0) = J_{\alpha\beta\gamma\delta}(L).
\end{equation}
The Cooperon $C$ and the propagator $\chi$ of Eqs.\ (\ref{cooperon}) and
(\ref{propagator}) are related to the density $N$ by 
\begin{eqnarray}
&& N_{\alpha\beta\gamma\delta} (x) = \int_0^\infty \! dt \, {\rm e}^{-t/\tau_\varphi}
\frac{1}{2\pi} \int_0^{2\pi} d\phi \int_0^{2\pi} d\phi_{\rm i} \,
\chi_{\alpha\beta\gamma\delta} (x,x_{\rm i};\phi,\phi_{\rm i};t), \\
&& \sum_{\alpha,\beta} N_{\alpha\beta\beta\alpha} (x_{\rm i}) =
\int_0^\infty \! dt \, {\rm e}^{-t/\tau_\varphi} C(t). 
\end{eqnarray}
Hence the weak-localization correction (\ref{wl1}) is obtained from $N$ by
\begin{equation}
\label{wl2}
\Delta G = -\frac{e^2D}{\pi\hbar L}
\sum_{\alpha,\beta} N_{\alpha\beta\beta\alpha}(x_{\rm i}).
\label{wlresult}
\end{equation}

The transformation to the local basis of spin states (\ref{localbasis}) takes
the form of a unitary transformation of the moments $N$ and $J$,
\begin{mathletters}
\begin{eqnarray}
&& \tilde{N}_{\alpha\beta\gamma\delta} = \sum_{\alpha',\beta',\gamma',\delta'}
Q^{\vphantom{\dagger}}_{\alpha\alpha'} Q^{\vphantom{\dagger}}_{\gamma\gamma'}
N^{\vphantom{\dagger}}_{\alpha'\beta'\gamma'\delta'}
 Q^\dagger_{\beta'\beta} Q^\dagger_{\delta'\delta}, \\
&& \tilde{J}_{\alpha\beta\gamma\delta} = \sum_{\alpha',\beta',\gamma',\delta'}
Q^{\vphantom{\dagger}}_{\alpha\alpha'} Q^{\vphantom{\dagger}}_{\gamma\gamma'}
J^{\vphantom{\dagger}}_{\alpha'\beta'\gamma'\delta'}
 Q^\dagger_{\beta'\beta} Q^\dagger_{\delta'\delta}, \\
&& Q(x)= 
\pmatrix{
\hphantom{-} {\rm e}^{i\pi fx/L}\,\cos\frac{\eta}{2}
& {\rm e}^{-i\pi fx/L} \, \sin\frac{\eta}{2} \cr
- {\rm e}^{i\pi fx/L} \, \sin\frac{\eta}{2}
& {\rm e}^{-i\pi fx/L} \, \cos\frac{\eta}{2} }.
\end{eqnarray}
\end{mathletters}
The transformed moments obey
\begin{mathletters}
\label{lodiff}
\begin{eqnarray}
\label{lcoopdiffa}
\ell\frac{d\tilde{N}_{\alpha\beta\gamma\delta}}{dx} &=&
 \sum_{\alpha',\gamma'}
\left(T_{\alpha\alpha'}\delta_{\gamma\gamma'}+\delta_{\alpha\alpha'}T_{\gamma\gamma'}\right)
\tilde{N}_{\alpha'\beta\gamma'\delta}
+\sum_{\alpha',\gamma'}
\left(S_{\alpha\alpha'}\delta_{\gamma\gamma'}-\delta_{\alpha\alpha'}S_{\gamma\gamma'}\right)
\tilde{J}_{\alpha'\beta\gamma'\delta} \nonumber \\
&& -\left(1+\tau/\tau_\varphi\right) \tilde{J}_{\alpha\beta\gamma\delta}, \\
\label{lcoopdiffb}
\ell\frac{d\tilde{J}_{\alpha\beta\gamma\delta}}{dx} &=& 2\sum_{\alpha',\gamma'}
\left(S_{\alpha\alpha'}\delta_{\gamma\gamma'}-
\delta_{\alpha\alpha'}S_{\gamma\gamma'}\right) \tilde{N}_{\alpha'\beta\gamma'\delta}
+\sum_{\alpha',\gamma'}
\left(T_{\alpha\alpha'}\delta_{\gamma\gamma'}+\delta_{\alpha\alpha'}T_{\gamma\gamma'}\right)
\tilde{J}_{\alpha'\beta\gamma'\delta} \nonumber \\
&&
-(2\tau/\tau_\varphi) \tilde{N}_{\alpha\beta\gamma\delta}
+ 2\tau \delta_{\alpha\beta} \delta_{\gamma\delta} \delta (x-x_{\rm i}),
\end{eqnarray}
\end{mathletters}
with the same $2\times 2$ matrices $S$ and $T$ as in Sec.\ \ref{transmission}.
Because the transformation from $N$ to $\tilde{N}$ is unitary, the weak-localization
correction is still given by 
$\Delta G=-(e^2D/\pi\hbar L)\sum_{\alpha,\beta}
\tilde{N}_{\alpha\beta\beta\alpha}(x_{\rm i})$, as in Eq.\ (\ref{wl2}).

\begin{figure}[ht]
\epsfxsize=0.7\hsize
 \hspace*{\fill}
 \vspace*{-0ex}\epsffile{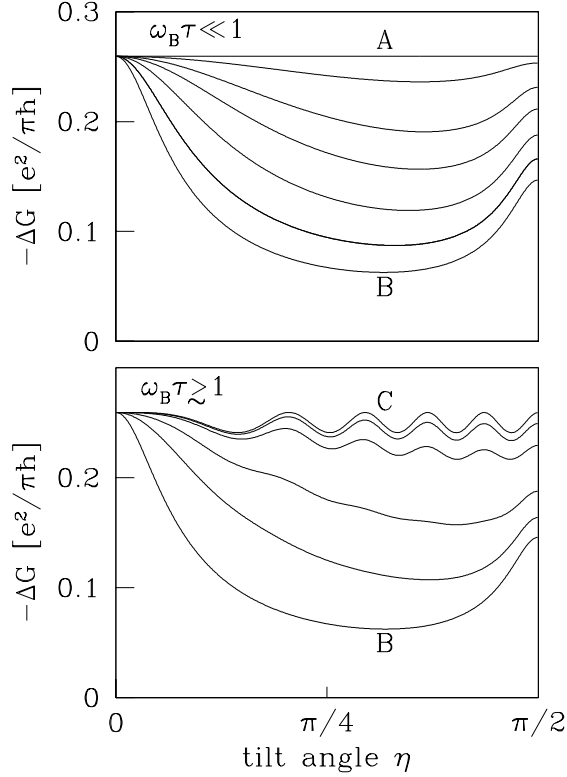}\vspace*{0ex}
 \hspace*{\fill}
\medskip
\caption[]{Weak-localization correction $\Delta G$ of a ring in a spatially rotating
magnetic field, as a function of the tilt angle $\eta$.
Plotted is the result of Eq.\ (\ref{lodiff}) for $f=5$, $L=500\,\ell$,
$L_\varphi=125\,\ell$.
The upper panel is for $\omega_{\rm B}\tau\ll 1$. From top to bottom:
$\omega_{\rm B}\tau = 10^{-5}$, $10^{-4}$, $2\cdot 10^{-4}$,
$3\cdot 10^{-4}$, $5 \cdot 10^{-4}$, $10^{-3}$, $10^{-2}$.
At $\omega_{\rm B}\tau\simeq (f\ell/L)^2$, the weak-localization correction crosses
over from the transient regime A
of Eq.\ (\ref{zerofield}) to the randomized regime B of Eq.\ (\ref{exactsol}).
The lower panel is for $\omega_{\rm B}\tau\gtrsim 1$. From bottom to top:
$\omega_{\rm B}\tau =0.1$, $1$, $2$, $5$, $10$, $100$.
Here the weak-localization correction reaches the adiabatic regime C
of Eq.\ (\ref{adiabatic}).
}
\label{fig3}
\end{figure}

We have solved Eq.\ (\ref{lodiff}) with periodic boundary conditions by numerically computing 
the eigenvalues and (right) eigenvectors of the $8\times 8$ matrix of coefficients.
The resulting $\Delta G$ is plotted in Fig.\ \ref{fig3} as a function of the tilt angle
$\eta$.
In the adiabatic regime $\omega_{\rm B}\tau\gg f$ we find the conductance oscillations
due to the Berry phase. These are given by \cite{lsg}
\begin{eqnarray}
\label{adiabatic}
&&\Delta G = -\frac{e^2}{\pi\hbar} \frac{L_\varphi}{L}
\frac{\sinh (L/L_\varphi)}{\cosh (L/L_\varphi)-\cos\left(2\pi f\cos\eta\right)}
\end{eqnarray}
analogously to the Aharonov-Bohm oscillations.\cite{ab}
(The length $L_\varphi =\sqrt{D\tau_\varphi}$ is the phase-coherence length.)
In the randomized regime $\left(f\ell/L\right)^2 \ll \omega_{\rm B}\tau \ll f$
there are no conductance oscillations. Instead we find a reduction of the
weak-localization correction, due to dephasing by spin scattering.
In the transient regime $\omega_{\rm B}\tau \ll \left(f\ell/L\right)^2$ the effect of
the field on the spin can be ignored,\cite{noot} and the weak-localization correction
remains at its zero-field value
\begin{eqnarray}
\label{zerofield}
&&\Delta G = -\frac{e^2}{\pi\hbar}\frac{L_\varphi}{L}\,{\rm cotanh}\left(\frac{L}{2L_\varphi}\right).
\end{eqnarray}

\subsection{Comparison with Loss, Schoeller, and Goldbart}
\label{sollsg}

If we replace the Boltzmann operator $\cal B$ in Eq.\ (\ref{kinetic}) by the diffusion operator
$-D\partial^2/\partial x^2$ and integrate over $\phi$ and $\phi_{\rm i}$,
we end up with the diffusion equation studied by LSG,
\begin{mathletters}
\label{lsgkinetic}
\begin{eqnarray}
&& \left(\frac{\partial}{\partial t}-{\cal H}\right)
\chi_{\alpha\beta\gamma\delta} (x,x_{\rm i};t) = \delta (t) \delta (x-x_{\rm i})
\delta_{\alpha\beta}\delta_{\gamma\delta}, \\
&& {\cal H} = D \frac{\partial^2}{\partial x^2}
+\frac{ig\mu_{\rm B}}{2\hbar}
\left[{\bf B}(x)\cdot \mbox{\boldmath$\sigma$}_1-
{\bf B}(x)\cdot \mbox{\boldmath$\sigma$}_2 \right], \\
&& \chi_{\alpha\beta\gamma\delta} (x,x_{\rm i};t) = \frac{1}{2\pi} \int_0^{2\pi}\! d\phi
\int_0^{2\pi}\! d\phi_{\rm i}\,
\chi_{\alpha\beta\gamma\delta} (x,x_{\rm i};\phi,\phi_{\rm i};t).
\end{eqnarray}
\end{mathletters}
Here $\mbox{\boldmath$\sigma$}_1$ and $\mbox{\boldmath$\sigma$}_2$ act, respectively, on
the first and third indices of $\chi_{\alpha\beta\gamma\delta}$.

The difference between the diffusion equation (\ref{lsgkinetic}) and the diffusion
equation (\ref{odiff}) is that (\ref{lsgkinetic}) holds only if $\omega_{\rm B}\tau\ll 1$,
while (\ref{odiff}) holds for any value of $\omega_{\rm B}\tau$.
LSG used Eq.\ (\ref{lsgkinetic}) to argue that there exists an adiabatic region
within the regime $\omega_{\rm B}\tau \ll 1$. In contrast, our analysis of 
Eq.\ (\ref{odiff}) shows that adiabaticity is not possible if $\omega_{\rm B}\tau\ll 1$.
The argument of LSG is based on a mapping of the diffusion equation
(\ref{lsgkinetic}) onto the Schr{\"o}dinger equation studied in Ref.\ \onlinecite{lg}.
However, the mapping is not
carried out explicitly. In this subsection we will solve Eq.\ (\ref{lsgkinetic})
exactly using this mapping, to demonstrate that the adiabatic regime of LSG is in fact
the randomized regime B. This mis-identification perhaps occurred because both
regimes are stationary with respect to the magnetic field strength
(cf.\ Fig.\ \ref{fig2}). However, Berry-phase oscillations of the conductance are
only supported in the adiabatic regime C, not in the randomized regime B
(cf.\ Fig.\ \ref{fig3}). 

We solve Eq.\ (\ref{lsgkinetic}) for the weak-localization correction
\begin{eqnarray}
 \label{wl3}
\Delta G &=& -\frac{e^2D}{\pi\hbar L} \sum_{\alpha,\beta}
\left\langle x,\alpha,\beta\left| \left(\tau_\varphi^{-1}-{\cal H}\right)^{-1}
\right|x,\beta,\alpha\right\rangle,
\end{eqnarray}
where we introduced the basis of eigenstates $|x,\alpha,\beta\rangle$
(with $\alpha,\beta=\pm 1$)
of the position operator $x$ and the spin operators $\sigma_{1z}$ and $\sigma_{2z}$.
The operator ${\cal H}$ commutes with
\begin{equation}
J=\frac{L}{2\pi i}\frac{\partial}{\partial x}+\case{1}{2}f\left(\sigma_{1z}+\sigma_{2z}\right).
\end{equation}
It is therefore convenient to use as a basis, instead of the eigenstates
$|x,\alpha,\beta\rangle$, the eigenstates $|j,\alpha,\beta\rangle$ of $J$,
$\sigma_{1z}$, and $\sigma_{2z}$. The eigenvalue $j$ of $J$ is an integer because of the
periodic boundary conditions. The eigenfunctions are given by
\begin{equation}
\left\langle x,\alpha',\beta'| j,\alpha,\beta\right\rangle = \frac{1}{\sqrt{L}}
\delta_{\alpha'\alpha} \delta_{\beta'\beta}
\exp\left[\case{2\pi i x}{L} (j-\case{1}{2}f\alpha-\case{1}{2}f\beta)\right] .
\end{equation}
In the basis $\{|j,1,1\rangle,|j,1,-1\rangle, |j,-1,1\rangle, |j,-1,-1\rangle \}$
the operator $\cal H$ has matrix elements
\begin{eqnarray}
\label{diagham}
\langle j',\alpha',\beta' | {\cal H} | j,\alpha,\beta\rangle &=&
-D \left(\frac{2\pi}{L}\right)^2 \delta_{j'j} 
\pmatrix{
(j-f)^2    &   0   &   0    &   0    \cr
0    &  j^2   &   0    &   0    \cr
0    &   0   & j^2    &   0    \cr
0    &   0   &   0    & (j+f)^2  \cr    
 }   \nonumber \\
&& -i\omega_{\rm B} \delta_{j'j}
\pmatrix{ 0  & \hphantom{-}\sin\eta  & -\sin\eta & 0 \cr
\hphantom{-}\sin\eta & -2\cos\eta & 0 & -\sin\eta \cr
-\sin\eta & 0 & 2\cos\eta & \hphantom{-}\sin\eta \cr
0 & -\sin\eta & \hphantom{-}\sin\eta & 0 \cr
}.
\end{eqnarray}

Substitution into Eq.\ (\ref{wl3}) yields
\begin{eqnarray}
\label{solwl}
\Delta G &=& -\frac{e^2D}{\pi\hbar}\frac{1}{L^2}\sum_{\alpha,\beta}
 \sum_{j=-\infty}^\infty
\left\langle j,\alpha,\beta \left| \left(\tau_\varphi^{-1} -{\cal H}\right)^{-1}
\right|j,\beta,\alpha \right\rangle \nonumber \\
&=& -\frac{e^2}{\pi\hbar}\frac{1}{2\pi^2}\sum_{j=-\infty}^\infty
\left[(\gamma + j^2)^2 (f^2 + \gamma + j^2) + 
      \kappa^2 (3 f^2 + 4 \gamma + 4 j^2 + f^2 \cos 2 \eta ) \right] \nonumber  \\
&& \mbox{}\times\left[(\gamma +j^2)^2 (f^4+2 f^2\gamma +\gamma^2-2 f^2j^2+2\gamma j^2 + 
       j^4) \right. \nonumber \\
&& \mbox{}+\left. 2 \kappa^2 \biglb( f^4 + 3 f^2 \gamma + 2 \gamma^2 - f^2 j^2 + 4 \gamma j^2 + 
       2 j^4 + f^2 (f^2 + \gamma - 3 j^2 ) \cos 2\eta \bigrb) \right]^{-1}.
\end{eqnarray}
We abbreviated $\kappa=2\omega_{\rm B}\tau (L/2\pi\ell)^2$ and $\gamma=(L/2\pi L_\varphi)^2$.
The sum over $j$ can be done analytically for $\kappa\gg 1$, with the result
\begin{mathletters}
\label{exactsol}
\begin{eqnarray}
&&\Delta G = -\frac{e^2}{\pi\hbar} \frac{1}{4\pi Q}
\left[
\frac{4a_-+4\gamma+(3+\cos 2\eta )f^2}{\sqrt{a_-}\tan\pi \sqrt{a_-}} -
\frac{4a_++4\gamma+(3+\cos 2\eta )f^2}{\sqrt{a_+}\tan\pi \sqrt{a_+}}
\right], \\
&& Q=\left[f^4 (9\cos^2 2\eta -2\cos 2\eta-7)-32\gamma f^2 (1+\cos 2\eta)\right]^{1/2}, \\
&&a_\pm = -\gamma + \case{1}{4}(1+3\cos 2 \eta ) f^2 \pm\case{1}{4}Q.
\end{eqnarray}
\end{mathletters}
We have checked that our solution (\ref{solwl}) of
Eq.\ (\ref{lsgkinetic}) coincides with the solution of Eq.\ (\ref{odiff})
in the regime $\omega_{\rm B}\tau\ll 1$.
(The two sets of curves are indistinguishable on the scale of Fig.\ \ref{fig3}.)
In particular, Eq.\ (\ref{exactsol}) coincides with the curves labeled B in
Fig.\ \ref{fig3}, demonstrating that it represents the randomized regime
-- without Berry phase-oscillations.

\section{Conclusions}

In conclusion, we have computed the effect of a non-uniform magnetic field on the
spin polarization (Sec.\ \ref{transmission}) and weak-localization correction
(Sec.\ \ref{localization}) in a disordered conductor. We have identified
three regimes of magnetic field strength: the transient regime
$\omega_{\rm B}\tau\ll (f\ell/L)^2$, the randomized regime
$(f\ell/L)^2 \ll \omega_{\rm B}\tau \ll f$, and the adiabatic regime
$\omega_{\rm B}\tau\gg f$. In the transient regime (labeled A in
Figs.\ \ref{fig2} and \ref{fig3}), the effect of the magnetic field can be
neglected. In the randomized regime (labeled B), the depolarization and the
suppression of the weak-localization correction are maximal. In the adiabatic
regime (labeled C), the polarization is restored and the weak-localization
correction exhibits oscillations due to the Berry phase. 

The criterion for adiabaticity is $\omega_{\rm B}t_{\rm c}\gg 1$, with $\omega_{\rm B}$
the spin-precession frequency and $t_{\rm c}$ a characteristic timescale of the orbital
motion. We find $t_{\rm c} =\tau$, in agreement with Stern,\cite{stern} but in
contradiction with the result $t_{\rm c}=\tau (L/\ell)^2$ of Loss, Schoeller,
and Goldbart. \cite{lsg} By solving exactly the diffusion equation for the Cooperon
derived in Ref.\ \onlinecite{lsg}, we have demonstrated unambiguously that the regime
which in that paper was identified as the adiabatic regime, is in fact the randomized
regime B --- without Berry-phase oscillations.

We have focused on transport properties, such as conductance and spin-resolved transmission.
Thermodynamic properties, such as the persistent current, in a non-uniform magnetic field
have been studied by Loss, Goldbart, and Balatsky \cite{lg,lgb} in connection with
Berry-phase oscillations. These papers assumed ballistic systems. We believe that the
adiabaticity criterion
$\omega_{\rm B}\tau\gg 1$ for disordered systems should apply to thermodynamic properties
as well as transport properties. This strong-field criterion presents
a pessimistic outlook for the prospect of experiments on the Berry phase in disordered 
systems.

\acknowledgements
We are indebted to L. P. Kouwenhoven for bringing this problem to our
attention, and to P. W. Brouwer, D. Loss, and A. Stern for useful discussions.
This research was supported by the ``Ne\-der\-land\-se or\-ga\-ni\-sa\-tie voor
We\-ten\-schap\-pe\-lijk On\-der\-zoek'' (NWO) and by the ``Stich\-ting voor
Fun\-da\-men\-teel On\-der\-zoek der Ma\-te\-rie'' (FOM).


\begin{references}
\bibitem{noot}
Because we include only the effect of the magnetic field on the spin, we do not
find the suppression of weak-localization due to time-reversal-symmetry breaking
of the orbital motion, nor do we find the Aharonov-Bohm oscillations due to the
coupling of the magnetic field to the charge.
\end{references}
\end{document}